\documentclass[aps,twocolumn,epsf,floats,prl,reprint,nofootinbib]{revtex4-1}
\usepackage{graphics,graphicx,epsfig}
\usepackage{amssymb,color}
\usepackage{epsf,epstopdf,wrapfig}
\usepackage{amsmath}
\usepackage{multirow}

\begin{document}

\title{Finite-size scaling as a way to probe near-criticality in natural swarms}

\author{Alessandro Attanasi$^{a,b}$, Andrea Cavagna$^{a,b,c}$, Lorenzo Del Castello $^{a,b}$, Irene Giardina$^{a,b,c}$,  Stefania Melillo$^{a,b}$, Leonardo Parisi$^{a,d}$, Oliver Pohl$^{a,b}$, Bruno Rossaro$^e$, Edward Shen$^{a,b}$, Edmondo Silvestri$^{a,f}$, Massimiliano Viale$^{a,b}$}

\affiliation{$^a$ Istituto Sistemi Complessi, Consiglio Nazionale delle Ricerche, UOS Sapienza, 00185 Rome, Italy}
\affiliation{$^b$ Dipartimento di Fisica, Universit\`a\ Sapienza, 00185 Rome, Italy}
\affiliation{$^c$ Initiative for the Theoretical Sciences, The Graduate Center, 365 Fifth Avenue, New York, NY 10016 USA}
\affiliation{$^d$ Dipartimento di Informatica, Universit\`a\ Sapienza, 00198 Rome, Italy}
\affiliation{$^e$ DeFENS, Universit\`a\ degli Studi di Milano, 20133 Milano, Italy}
\affiliation{$^f$ Dipartimento di Fisica, Universit\`a\ di Roma 3, 00146 Rome, Italy}

\begin{abstract}
Collective behaviour in biological systems is often accompanied by strong correlations. The question has therefore arisen of whether correlation is amplified by the vicinity to some critical point in the parameters space. Biological systems, though, are typically quite far from the thermodynamic limit, so that the value of the control parameter at which correlation and susceptibility peak depend on size. Hence, a system would need to readjust its control parameter according to its size in order to be maximally correlated. This readjustment, though, has never been observed experimentally. By gathering three-dimensional data on swarms of midges in the field we find that swarms tune their control parameter and size so as to maintain a scaling behaviour of the correlation function.  As a consequence, correlation length and susceptibility scale with the system's size and swarms exhibit a near-maximal degree of correlation at all sizes. 

\end{abstract}

\maketitle	

Intriguing evidence has been presented in the past few years suggesting that some biological systems are close to criticality, namely to a special point in the control parameters space characterized by unusually large correlation and susceptibility \cite{bialek_11}. 
Although reminiscent of self-organized criticality (SOC), this phenomenon is quite distinct, in that it does not appear to be as essentially dynamical as SOC, and it rather finds its natural description in terms of steady state ensemble distributions \cite{bialek_11}.
In all studies where the control parameter has been reported, though, its value has invariably been the result of inference through a model \cite{hastie}. Inference is potentially prone to the problem of data undersampling and therefore the alleged vicinity of the inferred control parameter to a critical point has been questioned \cite{marsili_11}.  Even though direct experimental measurements of long-range correlations and scaling laws provide inference-free evidence \cite{clauset+al_09, cavagna+al_10, mora+al_10}, one could still object that conservation laws plus off-equilibrium dynamics can produce long-range correlations generically, namely without the need to tune the control parameter \cite{grinstein_1990}. Therefore, the lack of a direct experimental measurement of the actual vicinity of the control parameter to its critical value is a major missing piece of evidence in the debate about criticality in biological systems.

To make things even more complicated, there cannot be just {\it one} critical value of the control parameter. The critical point is sharply defined only in the thermodynamic limit. However, all biological groups have finite size, $N$, which is often quite different from group to group. The only finite-size remnant of criticality is the peak of some susceptibility, whose position approaches the bulk critical point for large sizes \cite{amit,privman}. Thus, at finite size, the effective critical value of the control parameter depends on $N$.  A value of the control parameter that makes a small system `critical', will be quite off-critical for a much larger system, and {\it vice-versa}. For example, a very small Ising model at the bulk critical temperature is in fact deeply magnetized, with very small connected correlation. Hence, the parameters of a biological system cannot simply be tuned to their bulk critical value, as this value would not be `critical' at all for systems with small $N$. In order to observe critical behaviour, the control parameters must depend on the system's size. Therefore, in the discussion about criticality in biological systems we lack two crucial pieces of evidence: i) a direct experimental measurement of the control parameter (as opposed to model-based inference); 
ii) experimental evidence that in systems of different size $N$ the control parameter varies with $N$ in such a way to keep the system always close to the maximum of the susceptibility. The aim of this work is to address these two points.

\begin{figure}[t!]
\includegraphics[width=1.0\columnwidth]{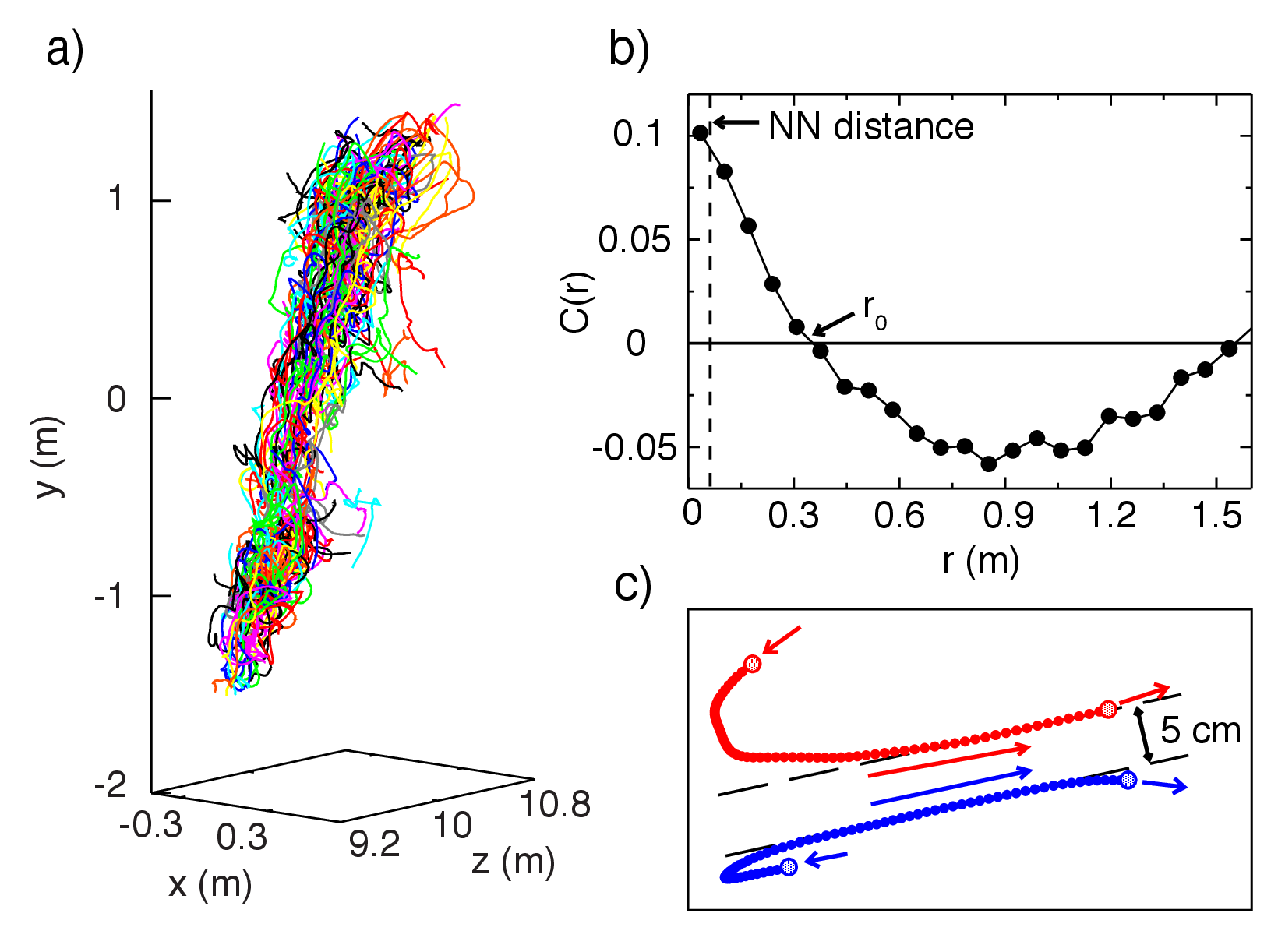}
\caption{{\bf a}. 3D trajectories for swarm $20120907\_$A$1$, $N=169$. {\bf b.} Velocity correlation function. The correlation length, $\xi\sim r_0$, is much larger than the nearest neighbour distance. The correlation is averaged over the whole time acquisition. {\bf c.} Alignment event between two midges (real trajectories).
}
\label{fig:correlation}
\end{figure}

We study wild swarms of midges in the field (Diptera: Chiro\-no\-midae  and Diptera: Cera\-topo\-gonidae) by reconstructing the $3d$ trajectories of individual insects within swarms ranging  from $100$ to $600$ individuals \cite{attanasi+al_13, attanasi+al_13b}. The $3d$ reconstruction of a swarm is shown in Fig.\ref{fig:correlation}a and in SM-Video 1. Swarms of diptera have been also studied in \cite{okubo_74,shinn+long_86,manoukis_09, ouellette+al_13}. Swarms are in a disordered phase, characterized by a low value of the alignment order parameter (average polarization, $\Phi = 0.2$ - see Table I in SM), but at the same time swarms exhibit significant directional correlations between individuals \cite{attanasi+al_13}. For each configuration, we define the equal-time, connected velocity correlation function as follows \cite{cavagna+al_10, attanasi+al_13}, 
\begin{equation}
C(r)=\frac{\sum_{i\neq j}^N \ \vec{\delta \varphi_i} \cdot \vec{\delta \varphi_j}\  \delta(r-r_{ij})}{\sum_{i\neq j}^N \  \delta(r-r_{ij})} \ ,
\label{corr}
\end{equation}
where $\delta\vec\varphi_i$ is the dimensionless velocity fluctuation, $\delta\vec{\varphi}_i=\delta \vec{v}_i/\sqrt{(1/N)\sum_k (\delta \vec{v}_k)^2}$,
and  $\delta \vec{v}_i$, is calculated by subtracting from the individual velocity $\vec{v}_i$ the contribution of the instantaneous global translation, rotation and dilatation of the swarm (see SM for details). The point where the correlation function first reaches zero, $C(r_0)=0$, is a finite-size proxy of the correlation length, $\xi$ (see SM). 
The integrated correlation,
\begin{equation}
\chi = \frac{1}{N} \sum_{i\neq j}^N \ \vec{\delta \varphi_i} \cdot \vec{\delta \varphi_j}\  \theta(r_0-r_{ij})  \ ,
\label{chi}
\end{equation}
is a finite-size proxy of the standard susceptibility computed from the fluctuations of the order parameter \cite{amit} (see SM) and for this reason we refer to it as the `susceptibility'. In a noninteracting system we find, on average, $\chi =0.1$ \cite{attanasi+al_13}.
In natural swarms $\chi\in [0.12:5.6]$ (see Table I in SM). Hence, the most correlated swarms have a susceptibility  over $50$ times larger than that of a noninteracting system. Large velocity correlations strongly suggest  that an effective alignment interaction is present in swarms. Indeed,  when two midges get closer than their metric interaction range (which is of the order of a few centimeters \cite{attanasi+al_13,fyodorova+al_03}) they tend to align their direction of motion (Fig.\ref{fig:correlation}c).

\begin{figure}[t!]
\includegraphics[width=1.0\columnwidth]{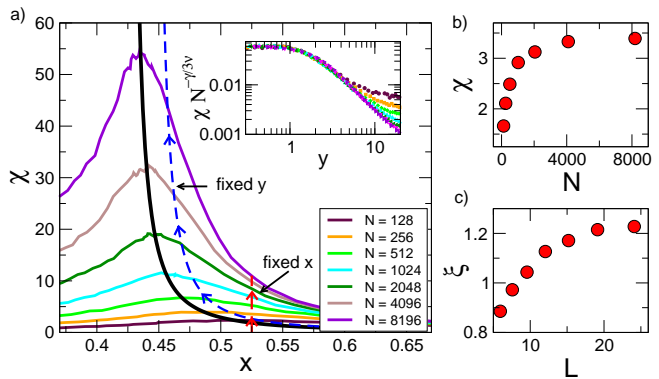}
\caption{
{\bf a.} 
Vicsek model in $3d$. Susceptibility $\chi$ as a function of the rescaled nearest neighbor distance, $x=r_1/\lambda$ for different swarm sizes, $N$. The maximum of $\chi$ occurs at the finite-size critical point, $x_\mathrm{max}(N)$, marked by the black line. 
Inset: rescaled susceptibility $\chi N^{-\gamma/3\nu}$ vs. scaling variable $y=(x-x_c) N^{1/3\nu}$. 
{\bf b.} Susceptibility as a function of  $N$ at fixed $x$. {\bf c.} Correlation length as a function of the linear system size, $L$, at fixed $x$.
By increasing $N$ (and $L$) at fixed value of $x$ we are moving along the red path in panel (a), so that 
we end up being further away from the the position of the maximum of $\chi$.  Simulations have been performed using the Vicsek update rule in $3d$: $ \vec{v}_i(t+1)=v_0 \ \mathcal{R}_\eta    (\sum_{r_{ij}<\lambda} \vec{v}_j(t))/|\sum_{r_{ij}<\lambda} \vec{v}_j(t))|$; $\vec{r}_i(t+1)=\vec{r}_i(t)+\vec{v}_i(t)$, where 
 $\mathcal{R}_\eta$ is a random uniform rotation in $[- 2\pi\eta, 2\pi \eta]$.  $v_0=0.05$, $\lambda=1$, $\eta=0.45$ (see SM).
}
\label{fig:vicsek}
\end{figure}

\begin{figure*}[t!]
\includegraphics[width=1.9\columnwidth]{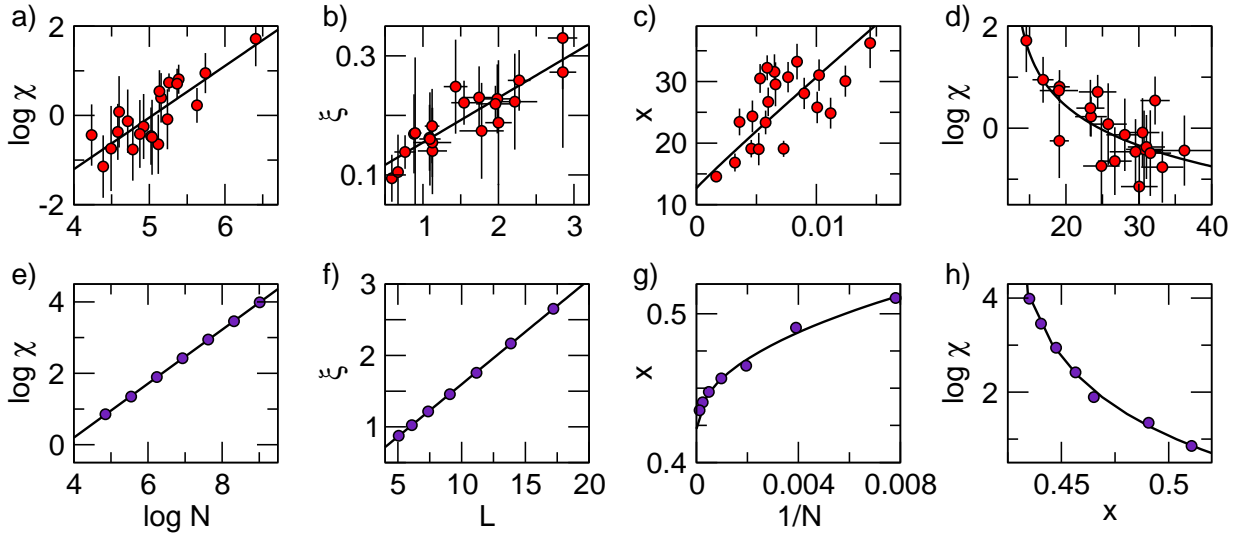}
\caption{
{\bf Top:} Natural swarms data. {\bf Bottom:} $3d$ Vicsek model in the critical region. {\bf a.} Susceptibility as a function of the number of midges, $N$ (P-value $= 3.0\times 10^{-6}$). {\bf b.} Correlation length, $\xi$, as a function of the linear system size, $L$ (P-value  $= 1.0\times 10^{-7}$). Both susceptibility and correlation length show no saturation for large systems. {\bf c.} Control parameter $x$  as a function of $N$ (P-value  $= 1.4\times 10^{-3}$).  {\bf d.} Susceptibility as a function of the control parameter, $x$ (P-value $= 6.9\times 10^{-5}$). Each point corresponds to a different swarm averaged over time (error bars are std deviations). Lower panels ({\bf e,f,g,h}): same quantities as in the upper panels, but calculated for the Vicsek model in the critical region, defined by a fixed value of the scaling variable $y$. This means that, unlike in Fig.2b,c, in panels {\bf e} and {\bf f} we are changing both $N$ {\it and} $x$ according to eq.(\ref{imma}) (blue path in Fig.2a).Lines are fit to eqs. (\ref{imma}-\ref{ummaz}). For $3d$ Vicsek we obtain: $\nu=0.75\pm0.02$, $\gamma=1.6\pm 0.1$, and $x_c=0.421\pm0.002$, not far from the $3d$ Heisenberg exponents \cite{pelissetto}. For natural swarms we obtain, $\nu = 0.35 \pm 0.1$, $\gamma= 0.9 \pm 0.2$, and $x_c = 12.5 \pm 1.0$. In natural swarms $L$ and $\xi$ are expressed in meters, while both $\chi$ and $x$ are dimensionless.
}
\label{fig:main}
\end{figure*}

Effective alignment, strong correlation and low order parameter are phenomena that finds a natural interpretation within Vicsek's model of collective motion \cite{vicsek+al_95}. In this model each individual aligns its velocity to that of neighbours within a metric interaction range, $\lambda$. At fixed low noise, the model exhibits a transition from a disordered phase (swarming) at low density, to an ordered phase (flocking) at high density. This density-driven transition is controlled by the parameter $x=r_1/\lambda$, namely the nearest neighbour distance, $r_1$, rescaled by the interaction range, $\lambda$. Hence, there is a value $x_c$ of the
control parameter below which spontaneous alignment emerges \cite{vicsek+al_95, gonci+al_08, chate+al_08}.
In the case of midges we do not know the interaction range $\lambda$. However, it was suggested in \cite{attanasi+al_13}
that the interaction between midges is acoustic, so that $\lambda$ is likely to be proportional to the body length, $l$. For this reason we can define
the control parameter of swarms as $x=r_1/l$ (see SM).

The bulk nature of the Vicsek transition is first order \cite{chate+al_08}; however, unless $N$ is very large, a pseudo second order phenomenology is observed, where all correlation markers (as $\xi$ and $\chi$) peak at the transition \cite{vicsek+al_95,baglietto+albano_08}. This ordering transition has been indeed observed in animal groups \cite{animal-transition}. Natural swarms of midges always exhibit low polarization and therefore live on the low-density, disordered side of the transition. Yet correlations are strong, suggesting that natural swarms are not too far from the transition. To investigate more precisely this point, though, we need a finite size scaling approach.

Finite-size scaling (FSS) has been studied in great details both in equilibrium \cite{amit,privman} and  in off-equilibrium  \cite{sides+al_98} systems. 
In the case of the Vicsek model a signature of the first-order nature of the transition occurs above a crossover size that is typically very large (e.g. $N\sim 10^6$ in $3d$, see \cite{chate+al_08} and SM). This means that below this size there exists a wide regime  (the one relevant for swarms) where FSS holds. This has been shown for the $2d$ Vicsek model in \cite{vicsek+al_95,baglietto+albano_08}.
Here we present evidence of FSS also in $3d$ (Fig.\ref{fig:vicsek}a): the susceptibility, $\chi$, has a peak at a pseudo-critical value $x_\mathrm{max}(N)$ of the control parameter, marking the finite-size crossover from a large $x$ disordered phase to a low $x$ ordered one. For larger $N$ the peak becomes sharper and shifts according to the FSS equation, $x_\mathrm{max}(N) = x_c + 1/N^{1/3\nu}$, where $\nu$ is the  critical exponent of the correlation length $\xi$ and $x_c$ is the bulk critical point. The scaling variable (at fixed noise) is thus, $y=(x-x_c)N^{1/3\nu}$, so that we expect susceptibility and correlation length to behave as, $\chi = N^{\gamma/3\nu} f(y)$ and $\xi = L \; g(y)$, where $f$ and $g$ are scaling functions. The scaling behaviour of the susceptibility in the $3d$ Vicsek model is quite satisfying (Fig.\ref{fig:vicsek}a, inset), giving $x_c=0.421\pm0.002$. Identical results hold in the more realistic case of a Vicsek model with harmonic confinement, which mimics the presence of the marker (see SM).

We can now use Fig.\ref{fig:vicsek}a as a map to interpret our expe\-ri\-men\-tal data. 
In the disordered phase, $x>x_c$, the rotational symmetry is unbroken (low polarization), hence no Goldstone mode is present \cite{amit} and the Vicsek model has a susceptibility and a correlation length which are finite in the infinite $N$ limit. Hence, by increasing $N$ at fixed $x$ (red path in Fig.2a), $\chi$ initially grows, but then it saturates to its finite bulk value for large $N$ (Fig.2b).
Consider two systems of sizes $L_1<L_2$, both of which are smaller than the {\it bulk} correlation length, $\xi_\infty$. When we increase the size of the group, passing from $L_1$ to $L_2$ all the individuals that we are adding are within a distance $\xi_\infty$ from each other and they are therefore strongly correlated; hence in this regime the finite-size $\xi$ grows with $L$ (Fig.\ref{fig:vicsek}c) and $\chi$ with $N$ (Fig.\ref{fig:vicsek}b). On the contrary, when $L>\xi_\infty$ an increase of the size amounts to adding particles statistical uncorrelated from each other, so that both $\xi$ and $\chi$ must saturate with the size (Fig.\ref{fig:vicsek}b,c).

In natural swarms, however, we do {\it not} observe a saturation of the susceptibility $\chi$, nor of the correlation length $\xi$, with the system's size. Instead, experimental data show that the susceptibility scales with $N$ and the correlation length scales with $L$ up to our largest sizes (Fig.~\ref{fig:main}a,b). There is nothing wrong with the aforementioned explanation, though: the saturation of $\chi$ and $\xi$ for large $N$ should only occur at {\it fixed} value of the control parameter, $x$.  Swarms, however, do not have a fixed value of $x$, but pick up their own values of $N$ and $x$. The fact that $\chi$ and $\xi$ show no hint of saturation suggests that when $N$ gets larger, $x$ decreases, as if swarms were following the peak of the susceptibility, yet remaining on the disordered side of the transition. This near-critical behaviour occurs when the control parameter $x$ and the system's size $N$ are related in such a way to keep constant the  scaling variable,  $y=(x-x_c)N^{1/3\nu}$, which is what happens along the blue path in Fig.2a. In this case, the following relations must hold,
\begin{eqnarray}
x & \sim& x_c + N^{-1/3\nu}  \ ,
\label{imma}
\\
\chi  &\sim& N^{\gamma/3\nu}  \ ,
\label{chinna}
\\
\xi &\sim& L   \ .
\label{xinna}
\end{eqnarray}
Equation (\ref{imma}) defines the near-critical region: it is this mutual readjustment of $x$ and $N$ that keeps the system scale-free, hence giving equations (\ref{chinna}) and (\ref{xinna}). Although the scatter is significant, the experimental data are compatible with equations (\ref{imma}-\ref{xinna}) (Fig.\ref{fig:main}a,b,c). In particular, we observe a correlation between control parameter $x$ and size $N$ (Fig.\ref{fig:main}c). This is the most prominent evidence that the data are in the near-critical region:  not only the correlation in swarms is scale-free ($\xi\sim L$, $\chi \sim N$), but a change in the size $N$ of the group is accompanied by a change in the control parameter $x$ as to compensate finite-size effects and {\it keep} the system scale-free correlated. If $(x,N)$ are in the near-critical region defined by (\ref{imma}),  the susceptibility must depend on $x$ as,
\begin{equation}
\chi \sim \frac{1}{(x - x_c)^\gamma}  \ ,
\label{ummaz}
\end{equation}
which is the black line in Fig.\ref{fig:vicsek}a.
Again, the scatter is large, but we can see from Fig.\ref{fig:main}d that the susceptibility of swarms indeed grows on decreasing the rescaled nearest neighbour distance $x$, with no evidence of a maximum, so that (\ref{ummaz}) does a fair job in fitting the data. In the lower panels of Fig.\ref{fig:main} we report the behaviour of  the $3d$ Vicsek model in the near-critical region, namely in the region defined by a constant value of the scaling variable $y=(x-x_c)N^{1/3\nu}$ (blue path in Fig.2a). The similarity with natural swarms is quite satisfying.

Even though we have data for smaller swarms ($N\ll 100$), we find that surface effects are too strong for these cases and that the statistical approach we use here is not justified anymore. On the other hand, at the moment it is technically hard to record swarms with $N\gg 10^3$. The span of our experimental data is therefore limited and different fits would work equally well. hence, the value of the critical exponents is far from conclusive (see also SM). Therefore, we simply claim that data are compatible with the FSS scenario of the Vicsek model and that the data show scaling. It is important to note that the result that natural swarms live in the near-critical region at the edge of an ordering transition is independent of the data fit.

What distinguishes our results about near-criticality from previous studies is that: i) we measure, rather than infer, the control parameter; ii) we do not simply find a generic vicinity of the control parameter to its bulk critical value, but we actually observe a mutual adjustment of control parameter and system's size that grants  the system  scale-free correlations.  This second result seems to rule out the `generic scale invariance' of \cite{grinstein_1990}. Note that when $N$ is rather small the pseudo-critical value of the control parameter, $x_\mathrm{max}(N)$, can be quite far from the bulk critical point, $x_c$. What matters is the balance between $N$ and $x$, not just the vicinity to $x_c$. When dealing with biological groups, where $N$ is never as large as in condensed matter, it is essential to keep in mind this finite-size scaling description of criticality. It is the pair $(x,N)$ that needs to be in the scaling region, not simply the control parameter.

There are two different ways of interpreting our results.
The first possibility is that, {\it given the size $N$}, the control parameter $x$ is tuned close to $x_\mathrm{max}(N)$, so that the group is endowed with large correlation. This mechanism requires individuals in the group to be able to assess global correlation by means of some local proxy, so that the control parameter $x$ can be readjusted if $N$ is varied. There is, however, an other interpretation. Instead of asking what is the optimal $x$ given $N$, we can ask what is the optimal $N$ given $x$. For each value $x$ of the control parameter, there is an optimal size $N_\mathrm{max}(x)$ (obtained by inverting equation (\ref{imma})) for which the maximum of the curve $\chi(x)$  occurs precisely at that $x$ (Fig.~\ref{fig:vicsek}a).
Hence, it is possible that {\it given the control parameter}, $x>x_c$, a group grows up to its maximum sustainable size, $N_\mathrm{max}(x)$.
For all values of $N<N_\mathrm{max}(x)$ the system is  in the ordered phase, where the correlation length scales with the system's size (due to Goldstone's mode). Hence the swarm can grow maintaining a constant level of relative correlation, $\xi/L $.
On the contrary, for $N>N_\mathrm{max}(x)$, the group would lose correlation with increasing size ($\xi/L\to 0$), leading to statistically independent clusters and a deterioration of collective response. Swarms have a mating purpose and male are naturally attracted to them \cite{downes_69}. Hence, an aggregation mechanism that leads to a maximum sustainable size is plausible. This might also explain why swarms do not order: the tendency to maximize the size of the group without decreasing correlation may drive the swarm away from the ordered phase, see also SM.

Scale-free correlations similar to those we have reported here for midges have been found in biological groups as diverse as bird flocks \cite{cavagna+al_10} and bacteria clusters \cite{chen}. Novel experiments trying to link correlation to collective response are needed to understand {\it why} correlation seems to be so widespread in biological systems.

{\bf Acknowledgments.}
We thank William Bialek, Yariv Kafri, Dov Levine and Victor Martin-Mayor for discussions. This work was supported by grants IIT--Seed Artswarm, ERC--StG n.257126 and US-AFOSR - FA95501010250 (through the University of Maryland).




\clearpage
\section*{Supplemental Material}
\section{Experiments}   

We performed stereoscopic experiments in the field (urban parks of Rome) between May and October, in $2011$ and in $2012$.
We acquired video sequences of natural swarms using a multi-camera system of three synchronized cameras (IDT-M5) shooting at $170$ fps. Two cameras (the stereometric pair) had a relative distance  $d$ in the interval 3-6m, depending on the swarm's distance and on the environmental constraints. A third camera, placed at a distance of $25\mathrm{cm}$ from the first camera, was used to solve tracking ambiguities. We used Schneider Xenoplan $50\mathrm{mm}$~$f/2.0$ lenses. Typical exposure parameters: aperture $f/5.6$, exposure time $3$ms. Recorded events have a time duration between $1.5$ and $15.8$ seconds. No artificial light was used. To reconstruct the $3d$ positions and velocities of individual midges we used the techniques developed in \cite{attanasi+al_13b}. 

In general, swarms  in the field form close to a na\-tu\-ral marker (a water puddle, some light foliage, etc). The marker is used by the midges within the  swarm to keep their absolute average position in space, possibly in order not to lose contact with the location where newly hatched females are \cite{downes_55}. The marker is therefore a source of stability for the swarm  (see \cite{attanasi+al_13b} for the role of the marker on the analysis of correlations). We found empirically that adding an extra artificial marker (a windscreen sun shield) on the grass beneath the swarm further increases the swarm's stability, hence giving us a longer time to mount the equipment and shoot the video sequence.

 After each acquisition we captured several midges in the recorded swarm for lab analysis. Midges were identified according to  \cite{Langton_07} (Chironomidae) and \cite{Kieffer_25} and \cite{Dominiak_12} (Ceratopogonidae). A summary of all swarms data can be found in Table 1.


\section{Connected correlation, susceptibility and scaling}

\subsection{The connected correlation function}

The connected correlation function $C(r)$ is usually defined by subtracting from the field the time (or ensemble) average.
In non-equilibrium systems with moving interaction network, like swarms are, we cannot do this and we must define the connected correlation using fluctuations with respect to instantaneous spatial averages. For a swarm at a given time $t$ the individual velocity is defined as, $\vec{v}_i(t)=[\vec{x}_i(t+\delta t)-\vec{x}_i(t)]/\delta t$ (or the sake of simplicity, in the rest of this Section we will set $\delta t=1$). We can compute the instantaneous average of the velocity over all the individuals $\vec{V}=(1/N) \sum_i  \vec{v}_i$, and then define the individual fluctuations with respect to this average $\delta \vec{v}_i=\vec{v}_i-\vec{V}$. These fluctuations measure how much the velocity of an individual midge differ from the instantaneous collective one. In systems where polar ordering is the only significant collective trait, these would certainly be the relevant fluctuations to look at. For swarms, however, we cannot a priori exclude that  other collective modes are present, as rotational or dilatational modes. If we want to monitor how individual behavior fluctuates with respect to the collective one, we must subtract these modes too, otherwise the connected correlation will not be correct. To do so, we proceed in the following way.
We first define the coordinate of each point in the center of mass reference frame, 
\begin{equation}
\vec y_i(t) =\vec x_i(t) - \frac{1}{N} \sum_k \vec x_k(t)  \ .
\end{equation}
Secondly, in the centre of mass reference frame we determine the optimal global rotation $\bf R$ and dilatation $\Lambda$ connecting two consecutive time frames, namely the rotation and dilatation that minimize
the quantity, $\sum_i [\vec{y}_i(t+1) - {\bf R}\Lambda\, \vec{y}_i(t)]^2$ \cite{attanasi+al_13}. 
The velocity fluctuation can then be defined as,
\begin{equation}
\delta \vec{v}_i=\vec y_i(t+1)-  {\bf R}\Lambda\, \vec{y}_i(t)  \ .
\end{equation}
Note that in the case where there is no rotation, nor dilatation, ${\bf R}=1, \Lambda =1$, we obtain,
\begin{equation}
\delta \vec{v}_i= 
\vec v_i(t)- \frac{1}{N} \sum_k \vec v_k(t) = \vec v_i(t)-\vec V
 \ ,
\end{equation}
which is the standard velocity fluctuation. On the other hand, if rotation and dilatation are nontrivial, then to each local velocity $\vec v_i$ we are subtracting the motion due to $\bf R$ and $\Lambda$ in that particular position.
The {\it dimensionless} velocity fluctuation appearing in the main text is defined as,
\begin{equation}
\delta\vec{\varphi}_i=\frac {\delta \vec{v}_i}{\sqrt{(1/N)\sum_k (\delta \vec{v}_k)^2}} \ .
\end{equation}
This definition is convenient because it allows to compare the fluctuations in system with widely different dimensional scales (for example, we cannot compare the velocity fluctuations in swarms measured in meters-per-seconds with those in a simulation).
The connected correlation is finally defined as,
\begin{equation}
C(r)=\frac{\sum_{i\neq j}^N \ \vec{\delta \varphi_i} \cdot \vec{\delta \varphi_j}\  \delta(r-r_{ij})}{\sum_{i\neq j}^N \  \delta(r-r_{ij})} \ .
\label{corr}
\end{equation}
Let us note that, by construction, we have $\sum_i \vec{\delta \varphi_i} = 0$. Therefore the correlation $C(r)$ as defined above must have a zero at a given point $r=r_0$, as can be seen in Fig.1 of the main text.

\subsection{The susceptibility}

The susceptibility $\chi$ is normally defined as the full volume integral of $C(r)$ and it measures the fluctuations of the global order parameter, i.e. $\rho \sigma^2 \int d^3r\ C(r) = N \langle \delta (\vec{V})^2 \rangle$ \cite{amit}. In swarms, however, we use definition (\ref{corr}), where fluctuations are considered with respect to space averages. In this case,  due to the constraint $\sum_i \vec{\delta \varphi_i} = 0$, the total volume integral is trivially equal to $-1$ and we cannot therefore use the standard definition of susceptibility. Moreover, calculating the susceptibility from the time fluctuations of the order parameter, as normally done in the literature \cite{baglietto+albano_08}, is unfortunately also not possible in swarms, as our experimental time series are not long enough (see below).
We therefore define the susceptibility as the maximum value reached by the integrated correlation, $\int^r d^3r' C(r')$. This maximum occurs when the correlation crosses zero, i.e. for $r=r_0$. We therefore define,
\begin{equation}
\chi = \rho\int^{r_0} d^3r \; C(r)  \ ,
\label{zona}
\end{equation}
where $\rho$ is the density. The integral in (\ref{zona}) gives an estimate of the volume of the correlated regions, so that $\chi$ is proportional to the number of correlated individuals in the system. If we make the hypothesis (experimentally verified) that mass fluctuations are not strong, we can write
$\sum_{ij}  \delta(r-r_{ij}) \sim 4\pi r^2 N\rho$, and obtain from (\ref{corr}) and (\ref{zona}) a binning-free definition of the susceptibility,
\begin{equation}
\chi = \frac{1}{N} \sum_{i\neq j}^N \ \vec{\delta \varphi_i} \cdot \vec{\delta \varphi_j}\  \theta(r_0-r_{ij})  \ ,
\label{chi}
\end{equation}
which we used in the main text. 

\subsection{Relationship with the standard susceptibility}

\begin{figure}[t!]
\includegraphics[width=0.9\columnwidth]{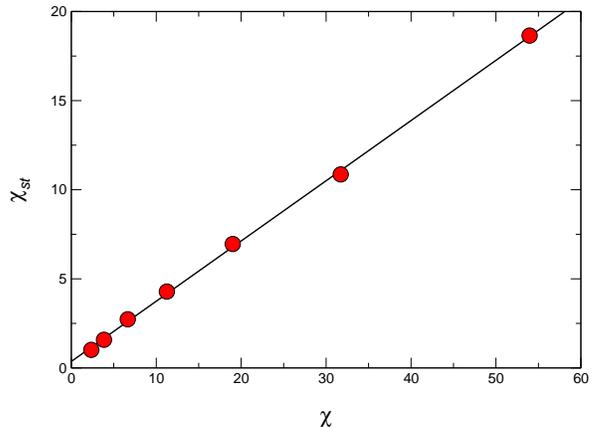}
\caption{
Viscek model in $d=3$. 
Standard susceptibility from the fluctuations of the order parameter, $\chi_\mathrm{st}$ (defined in (\ref{standard})) vs. susceptibility from the integrated correlation function, $\chi$ (defined in (\ref{chi})). Every point corresponds to different values of $x$ and $N$, in the scaling region of constant $y$, corresponding to the maximum of the susceptibility.}
\label{fig:chi}
\end{figure}

In systems where one has long enough time series that it is possible to compute time averages, the standard susceptibility, $\chi_\mathrm{st}$, is computed from the fluctuations of the order parameter \cite{baglietto+albano_08},
\begin{equation}
\chi_\mathrm{st} = \frac{N}{\sigma^2} \left(\langle \vert \vec V \vert^2\rangle - \langle \vert \vec V \vert \rangle^2 \right)  \ ,
\label{standard}
\end{equation}
where $\vec V =1/N \sum \vec v_i $ and
$\sigma^2=(1/N)\sum_i ( \vec{v}_i-\vec V)^2$. The brackets $\langle \cdots\rangle$ indicate averages over time. As we said, we do not have long enough time series to measure $\chi_\mathrm{st}$ and this is why we use (\ref{chi}). However, these two quantities have the same scaling behaviour.
In Fig.~\ref{fig:chi} we report for the Vicsek model in $d=3$ the susceptibility $\chi$ vs. $\chi_\mathrm{st}$, for different values of $x$ and $N$, in the scaling region, i.e. at fixed value of the scaling variable $y=(x-x_c)N^{1/3\nu}$. We clearly see that the two definitions of the susceptibility are simply proportional to each other in the scaling region. 

In equilibrium systems the susceptibility $\chi$ is proportional to the collective response of the system, i.e. to the derivative of the order parameter with respect to an external field (static fluctuation dissipation theorem). In off-equilibrium systems as swarms and flocks, we cannot prove such relation. Although it is somewhat natural to expect that in general the response is related to the amount of correlation in the system, new experimental data are needed to quantify this expectations. From the numerical point of view, an investigation of this point in the Vicsek model can be found in \cite{chate+al_08}.

\subsection{Scaling relations.}

We provide here some basic derivations of the finite-size scaling relations that we use in the main text. Reference \cite{amit} can be used 
as excellent background reading.

As we wrote in the main text, we use $r_0$ as a proxy for the correlation length, $\xi$. Let us see that this definition makes sense.
In a system with finite size $L$, we have,
\begin{equation}
C(r) = G(r) - \frac{3}{L^3} \int_0^L dr \; r^2 \, G(r) \ ,
\end{equation}
where $G(r)$ is the {\it bulk} correlation function, i.e. the correlation function in an infinitely large system. 
The equation, $C(r_0)=0$, therefore gives,
\begin{equation}
G(r_0) = \frac{3}{L^3} \int_0^L dr \; r^2 \, G(r) \ ,
\label{errezero}
\end{equation}
In the scaling region, we have that the correlation function has a scale-free form,
\begin{equation}
G(r)=  \left(\frac{\lambda}{r}\right)^{1+\eta} \quad , \quad d=3   \ ,
\label{bulk}
\end{equation}
where $\eta$ is the anomalous dimension and $\lambda$ is the range of the interaction, making the correlation function dimensionless.
By plugging (\ref{bulk}) into (\ref{errezero}),  we obtain that in the scaling region the correlation length scales with the system's linear size,
\begin{equation}
\xi \sim r_0 \sim L  \ ,
\label{sf}
\end{equation}
as expected. By using this last equation into (\ref{zona}) we finally obtain the susceptibility in the scaling region, 
\begin{equation}
\chi \sim (1/r_1)^3 \lambda^{1+\eta} L^{2-\eta}
  \ , 
\label{bonga}
\end{equation}
By using the scaling relation, $2-\eta = \gamma/\nu$, and by exploiting the equation, $L \sim r_1 N^{1/3}$, we obtain,
\begin{equation}
\chi \sim \frac{1}{x^{3-\gamma/\nu}} \, N^{\gamma/3\nu}
  \ , 
\label{bongo}
\end{equation}
where, as in the main text, we have defined $x\equiv r_1/\lambda$.
In a system with topological interaction everything must be invariant under rescaling of the nearest neighbour distance $r_1$, hence $\lambda\sim r_1$, as it happens in bird flocks \cite{ballerini+al_08a} and so the prefactor in (\ref{bongo}) is of order $1$. In this case (\ref{bongo}) is equivalent to the standard finite size scaling relation,
\begin{equation}
\chi(N) \sim N^{\gamma/3\nu}
  \ . 
\label{budu}
\end{equation}
On the other hand, in a metric system,  $r_c$ does not scale with $r_1$, hence the prefactor $x^{-(3-\gamma/\nu)}$ in (\ref{bongo}) remains. However, in our data the exponent $3-\gamma/\nu$ is very small, hence this correction to standard scaling is small.

\section{The Vicsek model}
\subsection{The standard Vicsek model in 3$d$}
\begin{figure}[b!]
\includegraphics[width=0.9\columnwidth]{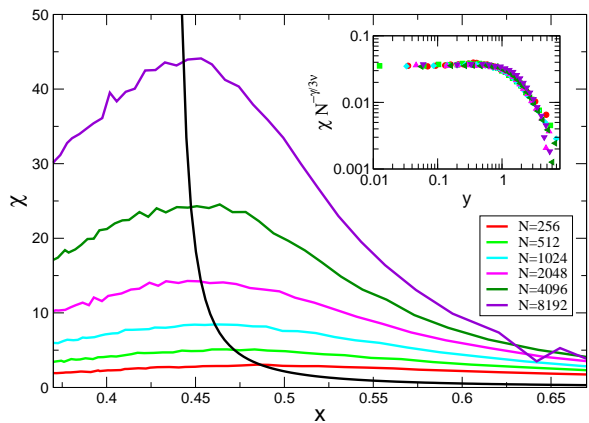}
\caption{
{\bf Finite size scaling of the Vicsek model in an harmonic potential.} 
Susceptibility $\chi$ as a function of the rescaled nearest neighbor distance, $x=r_1/\lambda$ for different swarm sizes $N$. The maximum of $\chi$ occurs at the finite-size critical point, $x_\mathrm{max}(N)$. This maximum becomes sharper and sharper for increasing $N$. The black line marks the critical line $x_\mathrm{max}(N)$.
Inset: rescaled susceptibility $\chi N^{-\gamma/\nu}$ vs. scaling variable $y=(x-x_c) N^{1/3\nu}$ for $x>x_c$. 
$v_0=0.05$, $\lambda=1$, $\eta=0.45$.
}
\label{fig:vicsek2}
\end{figure}

\begin{figure*}[t!]
\includegraphics[width=1.8\columnwidth]{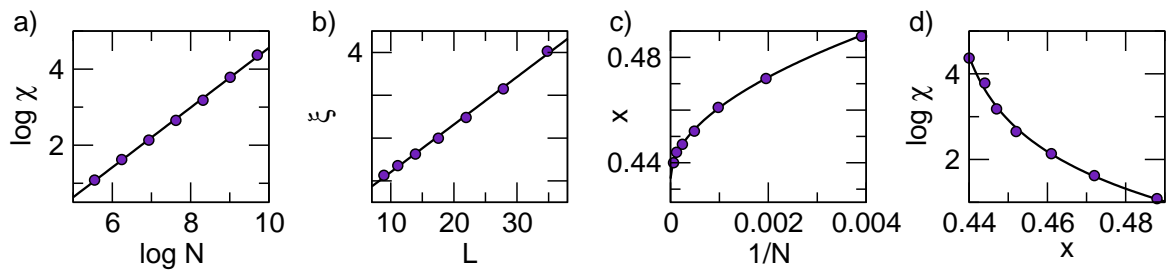}
\caption{
{\bf Scaling behavior in the Vicsek model with harmonic force.}
a) Susceptibility as a function of the number of midges $N$ in the swarm. b) Correlation length, $\xi$, as a function of the linear system size, $L$. c) Control parameter $x$  as a function of system size $N$ d) Susceptibility as a function of the control parameter $x$. Each point corresponds to a pair $(N,x_\mathrm{max}(N))$ along the critical line. 
 }
\label{fig:main2}
\end{figure*}
We performed numerical simulations of the Vicsek model in $3d$ \cite{vicsek+al_95,czirok+al_99,gonci+al_08,baglietto+albano_08,chate+al_08}. The direction of particle $i$ at time $t+1$ is the average direction of all particles within a sphere of radius $\lambda$ around $i$ (including $i$ itself). The parameter $\lambda$ is the metric radius of interaction. The resulting direction of motion is then perturbed with a random rotation (noise).  The update equations of this model read:
\begin{eqnarray}
\label{viccent_array}
 \vec{v}_i(t+1)& =& v_0 \ \mathcal{R}_\eta \left[  \Theta \left( \sum_{j\in S_i} \vec{v}_j(t) \right ) \right] \ ,\\
 \vec{x}_i(t+1)&=& \vec{x}_i(t)+\vec{v}_i(t+1) \ ,
\end{eqnarray}
where $v_0$ is the (fixed) speed of the particle, $S_i$ is the spherical neighborhood of radius $\lambda$ centered around $i$,  $\Theta$ is the normalization operator, $\Theta(\vec{x})=\vec{x}/|\vec{x}|$, and $\mathcal{R}_\eta$ performs a random rotation uniformly distributed around the argument vector with maximum amplitude of $4\pi\eta$.

We run simulations of this model with periodic boundary conditions, for several densities $\rho$ and systems sizes $N$.   Each simulation has a duration  of $6\times10^5$ time steps, with initial conditions consisting in uniformly distributed positions and uniformly distributed directions in the $4 \pi$ solid angle. After a transient of $10^5$ time steps, we saved 500 configurations at intervals of 1000 time steps in order to have configurations with velocity fluctuations uncorrelated in time. 

The Vicsek model exhibits a transition from a disordered phase (at large noise/low density) to a polarized `flocking' phase (low noise/large density). There is in fact a critical line in the $\eta,\rho$ plane characterizing this transition. In most animal groups the order-disorder transition described by the Vicsek model has been observed with respect to density \cite{buhl+al_06,becco_02,makris+al_09}.  For this reason, we consider the model with fixed noise value and focus on the behavior of the system with varying the density. We use as control parameter the average nearest-neighbor distance rescaled by the interaction range $x \equiv r_1/\lambda$. The average nearest-neighbor distance is in fact a measure of density ($r_1\sim \rho^{-1/d}$). Usually in numerical simulations the interaction range $\lambda$ is set to $1$, however this is not generally the case in nature. The reason why density is important to determine the macroscopic properties of the system is that the larger the density, the larger the number of neighbors within the interaction range, the stronger local ordering. Thus, what is relevant is how dense is the system with respect to the interaction range. For this reason we use $x$ as control parameter, rather than simply $r_1$ (or $\rho$).

The nature of the ordering transition in the Vicsek model has been the issue of a long debate. Recent analysis \cite{chate+al_08} indicate that the transition is first-order in the bulk. There are however very strong finite size effects, and this is why many works (e.g. \cite{vicsek+al_95, czirok+al_99,gonci+al_08}) show instead a second-order like phenomenology.  In particular, a signature of the first order transition only occurs at very large sizes $N>N^\star$, when the Binder cumulant develops a drop at negative values. The crossover size $N^\star$ depends on the parameters of the model (e.g. the density and the particles speed) and the kind of noise (scalar vs. vectorial), but it is in general very large, especially for the scalar noise case. For example, in $3d$  and for the parameters in \cite{chate+al_08}, one has  $N^\star\sim 10^6$. This means that there is a  wide regime where the phenomenology of the model is consistent with a continuos second order transition \cite{nagy+al_06} and where standard Finite Size Scaling (FSS) can therefore be used to describe the behavior of the system, and how it changes with size. Swarms and animal groups usually have moderate sizes, much smaller than $N^\star$. For this reason we are interested in the regime $N< N^\star$ where we can expect (and we in fact find) FSS to hold.

To understand the behavior of the system at finite size we therefore applied standard FSS \cite{amit,privman,sides+al_98,baglietto+albano_08}. At each fixed value of the system's size $N\in[128,8192]$ we calculated $\chi(x;N)$, as displayed in Fig.2 of the main text. To compute the correlation length we computed the connected velocity correlation $C(r)$ (see main text), and evaluated $\xi$ as
\begin{equation}
\xi = \frac{\int dr \ r^2 C(r)}{\int dr\ r C(r) } \ .
\label{len}
\end{equation}
This definition is equivalent to the alternative one where $\xi$ is identified with the zero of the correlation ($r_0$ - see main text) in the region where the correlation is long range, and is more appropriate in the deeply  disordered phase where the correlation decays exponentially.

To investigate the behavior in the critical region, we worked out the maximum of the susceptibility $\chi_\mathrm{max}(N)$ and its position $x_\mathrm{max}(N)$; these quantities as a function of $N$ are plotted in Fig.~3e and Fig.~3g in the main text. We obtained Fig.~3f by evaluating the correlation length (\ref{len}) along the critical line (i.e. for all values of the pair $(N,x_\mathrm{max}(N))$. Finally, we plotted $\chi_\mathrm{max}$ vs. $x_\mathrm{max}$ parametrically in $N$, to obtain the function $\chi(x)$ in Fig.~3h.  
All the above curves can be fitted using the predictions of FSS theory \cite{amit,privman,baglietto+albano_08}. In this way we estimated the exponents and the bulk value of the critical point $x_c$ (see main text).

 The noise, $\eta$, affects the height of the susceptibility peak and the position of the transition point  $x_c$ \cite{vicsek+al_95,gonci+al_08,chate+al_08}, but this is irrelevant for us, because we do not use  any quantitative result from the model to infer any biological parameters of real swarms. 
The data reported in the main text have $\eta=0.45$.

\subsection{The Vicsek model in an harmonic potential}

In the Vicsek model the particles are only subject to the `social' alignment force due to neighbors and the system is fully translationally invariant in space. 
Natural swarms, however, are known to  form close to a marker and to keep a stationary position with respect to it \cite{downes_55}. 
To mimic this behavior we can easily modify the Vicsek model by adding an external harmonic force equal for all particles. This potential also grants cohesion, without the need to introduce an inter-individual attraction force \cite{okubo_86,ouellette+al_13,butail+al_13}.
The update equation for velocities is in this case given by,
\begin{equation}
\label{viccent}
 \vec{v}_i(t+1)=v_0 \ \mathcal{R}_\eta \left[  \Theta \left( \sum_{j\in S_i} \vec{v}_j(t) - \beta\vec{r}_i(t)\right) \right] \ ,
\end{equation}
where $\beta$ is a parameter modulating the strength of the central force.

We investigated this variation of the Vicsek model (in $d=3$) with the same protocol described above for the standard Vicsek case. Now, thanks to the central force, we can use open boundary conditions (while in standard Vicsek this would lead to a dispersion of the group \cite{vicsek+al_95}). The density of the flock (and therefore $r_1$ and $x$) can be tuned by changing $\beta$, which sets the confinement volume for the swarm.  Also this modified Vicsek model  has a density driven transition from a disordered state to an ordered one. In the standard Vicsek the ordered phase consists of a polar flow of particles moving straight in the same direction \cite{vicsek+al_95}; in this modified version it corresponds to a coherent polarized flock orbiting around the centre of the harmonic potential (this is however not relevant  for our analysis, as natural swarms live in the disordered phase).

The FSS properties and the critical behavior of this modified Vicsek model are very similar to the standard $3d$ Vicsek model  presented in the main text, as can be seen in Fig.~\ref{fig:vicsek2} and Fig.~\ref{fig:main2} (the analogues of Fig.~2 and Fig.~3 in main text). The value of the exponents and of the bulk critical point are also very similar. We get 
$\nu=0.74 \pm 0.05$, $\gamma=1.5 \pm 0.2$, and $x_c=0.433 \pm 0.002$ (to be compared with $\nu=0.75\pm0.02$, $\gamma=1.6\pm 0.1$, and $x_c=0.421\pm0.002$ of the standard Vicsek case).

We add here a remark for future research.
As we have seen, the stiffness $\beta$ of the harmonic force regulates the size of the swarm, so one could think of a simulation where this stiffness is changed. In particular, when the harmonic constant of the potential gets smaller, not only the swarm gets larger (lower density), but the harmonic potential also becomes flatter (lower second derivative), hence more prone to external perturbations. It would then be interesting to study how the response of the `swarm' to external perturbations depends on this stiffness, and eventually check to what extent large correlation can help the swarm to keep steady and cohesive in an increasingly 
flat/unconfining potential.

\subsection{About the critical exponents}
Concerning the critical exponents, we note that the values we obtained from fitting the experimental data (see main text) are different from the ones obtained from numerical simulations for the $3d$ Vicsek model, with or without harmonic potential. On one hand, the span of our data is not very large and the precise values of the fitted exponents cannot, therefore, be fully trusted. On the other hand, there are several reasons why there could in fact be a difference.

One is dimensionality: the critical exponents generally depend both on the dimension $d$ of space and on that of the order parameter, $D$. Natural swarms live surely in $d=3$, but because of gravity there may be an effective dimensional reduction of $D$, similar to what happens in bird flocks: if the animal tends to save energy it will mostly fly level (small vertical displacement), hence effectively reducing $D$. An example of how this reduction can affect the exponents is given in \cite{toner+tu_98} (see the section dedicated to the anisotropic easy plane case). This factor has not been taken into account in our Vicsek simulations. Another possible source of difference are inertial effects: Vicsek is fully dissipative, whereas non-dissipative inertial terms could be present; how/whether these terms could change the exponents is unclear, but we cannot exclude this effect.

Finally, there is symmetry: this is one of the most relevant factors for critical exponents; as we mentioned, real swarming happens under gravity, which is a symmetry breaking direction. This argument is connected, although not identical, to the effective dimensional reduction of $D$. 
In principle, we could try to generalize the Vicsek model by adding a symmetry breaking term to mimic gravity, and second order derivatives to model inertial effects. However, our intention in this paper is not to reproduce in a detailed way the behavior of the swarms, nor to make strong claims on their dynamical universality class. Rather, we focus on the very general scaling behavior exhibited by the correlation and the susceptibility and how it can be interpreted as a signature of finite-size criticality. For this reason, we studied numerically the simplest possible model where these features are present and can be exhaustively characterized.

Note that the measurement of $\chi(x)$ and of $x(N)$ has been made possible by the fact that the interaction in  swarms is {\it metric} \cite{attanasi+al_13}, so that density is (through $x$) the control parameter.  In bird flocks, on the contrary, density seems to be irrelevant, due to the {\it topological} nature of the interaction  \cite{ballerini+al_08a}, whereas the control parameter is not directly measurable in experiments \cite{PNAS2014}. Hence, the present FSS analysis cannot be performed in flocks

\section{The control parameter in natural swarms}
As we have seen, the correct control parameter for the density-driven transition in a Vicsek-like system is the nearest neighbour distance rescaled by the interaction range, 
$x=r_1/\lambda$. We can easily measure $r_1$ in natural swarms, but we do not have an {\it a priori} knowledge of the interaction range, $\lambda$. This would not be a problem if our data were only from a single species, as we could reasonably assume $\lambda$ to be approximately the same within the same species. However, we have data from three different species (see Table I) and it would be a waste not to be able to use all the data together in our scaling plots. If we use different species, though, we can no longer assume that $\lambda$ is the same for all, hence we have to redefine the control parameter $x$ in some way.

This issue was studied in \cite{attanasi+al_13}, where it was hypothesized (by following a simple scaling argument) that each species is characterized by one single length-scale, namely its body length, $l$. If this is the case, then the interaction range will be proportional to $l$, so that using $x=r_1/l$ as a control parameter is equivalent to using $x=r_1/\lambda$. The clear advantage of using $l$ rather than $\lambda$ is that the body length can be actually measured for the midges involved in our study.

To justify the hypothesis that $\lambda \propto l$ we have two arguments. First, the susceptibility $\chi$ really seems to be a natural function of $r_1/l$ rather than simply $r_1$: the P-value of $\chi(r_1)$ is 0.07, whereas the P-value of $\chi(r_1/l)$ is 0.00007, namely an increase of three orders of magnitude in the statistical significance of the correlation between susceptibility and control parameter \cite{attanasi+al_13}. Even though this is a rather {\it a posteriori} motivation, it is quite a compelling one notwithstanding. The second argument is biological. We find in \cite{attanasi+al_13} that the interaction between midges is metric, and that its range is compatible with an acoustic interaction: midges perceive the wing flapping of other individuals. This thesis is also supported by the experiments described in \cite{McKie_Crabston_2005}, revealing the importance for midges to swarm of the acoustic perception through the antennae and the Johnston's organ. As shown in \cite{fedorova_2009}, both the sizes of the wings and of the antennae  are proportional to the body length. Hence, given that all physiological length scales involved in the interaction are proportional to $l$, it seems reasonably consistent to conclude that the length scale of the interaction too is proportional to it, namely $\lambda \propto l$. This is not a proof, of course, but together with the aforementioned increase in statistical significance when using $r_1/l$, it suggests that what we are doing is reasonable.

Of course, the best thing to do would be to rescale $r_1$ by the real interaction range. As a matter of fact, in \cite{attanasi+al_13} we managed to give an estimate of the (metric) interaction range $\lambda$ (it turns out that $\lambda$ is about 2-5cm, compatible with an acoustic-auditory interaction).
However, this very estimates of $\lambda$ uses the assumption that $r_1/l$ is the right scaling variable. Hence, using this estimate of $\lambda$ to rescale $r_1$ would be rather circular. In absence of an independent determination of $\lambda$, the best we can do is to use $x=r_1/\lambda$ as a control parameter.


\section{Continuous vs discrete symmetry breaking}
When looking for an explanation of near-criticality, we suggest in the main text that instead of being the control parameter $x$ adapting to the size $N$, it may be $N$ that grows up to the maximum sustainable size, given $x$. The idea is that all $N$ smaller than this maximum sustainable size $N_\mathrm{max}$ sustain scale-free correlations, while for $N > N_\mathrm{max}$ the correlation length $\xi$ saturates, so that $\xi/L$ starts decreasing. Beyond this point, larger sizes of the swarm becomes counterproductive.

This aggregation way to near-criticality relies on the fact that the ordered phase is characterized by a {\it continuous} symmetry breaking (the rotational symmetry in our case), so that the bulk susceptibility and correlation length are infinite in the ordered phase (Goldstone mode). In the case of a {\it discrete} symmetry (as for the Ising variables used in neural systems) the aggregation mechanism we propose would not work: if $N < N_\mathrm{max}(x)$, the (connected) correlation length does not grow as fast as the system's size, so that a fully correlated group is achieved only for $N\sim N_\mathrm{max}(x)$, not for lower, nor for larger groups. We conclude that in the
case of discrete symmetry breaking an adaptive mechanism of the control parameter $x$, given the size $N$, seems to be required to explain near-critical data.



\begin{table*}[t!]
\vskip 0.1 in
\begin{tabular}{l|c|c|c|c|c|c|c|c|c}
\hline
\hline
{\sc Species}   &\hspace{0.2cm}
 {\sc Event  label}   \hspace{0.2cm}     & \hspace{0.1cm}
 $N$    \hspace{0.1cm}     & \hspace{0.1cm}
 {\sc Duration (s)}    \hspace{0.1cm}     & \hspace{0.1cm}
 $l$ (mm)    \hspace{0.2cm}     & \hspace{0.1cm}
 $r_1$ (m)    \hspace{0.2cm}     & \hspace{0.1cm}
 $r_0$ (m)    \hspace{0.2cm}     & \hspace{0.1cm}
 $|\vec{v}|$ (m/s)    \hspace{0.2cm}     & \hspace{0.1cm}
 $\chi$    \hspace{0.2cm}     & \hspace{0.1cm}
 $\phi$
 \\
\hline
\hline
\multirow{3}{*}{\begin{minipage}[c]{3.5cm}\it Corynoneura scutellata\\ {\rm (Diptera: Chironomidae)} \\  \end{minipage}}
& 20110906\_A3 & 138 & 2.0  & 1.5 & 0.029 & 0.094 & 0.12 & 0.78 & 0.17 \\
& 20110908\_A1 & 119 & 4.4  & 1.1 & 0.036 & 0.105 & 0.13 & 0.46 & 0.27 \\
& 20110909\_A3 & 312 & 2.7  & 1.5 & 0.026 & 0.138 & 0.12 & 2.58 & 0.22 \\
\hline
\multirow{8}{*}{\begin{minipage}[c]{3.5cm}\it Cladotanytarsus  atridorsum \\ \rm  (Diptera: Chironomidae) \\  \end{minipage}}
& 20110930\_A1 & 173 & 5.9  & 2.4 & 0.057 & 0.228 & 0.23 & 1.48 & 0.31 \\
& 20110930\_A2 &  99 & 5.9  & 2.4 & 0.063 & 0.223 & 0.15 & 1.08 & 0.20 \\
& 20111011\_A1 & 131 & 5.9  & 2.4 & 0.075 & 0.272 & 0.11 & 0.65 & 0.17 \\
& 20120828\_A1 &  89 & 6.3  & 2.5 & 0.062 & 0.188 & 0.17 & 0.48 & 0.22 \\
& 20120907\_A1 & 169 & 3.2  & 1.9 & 0.062 & 0.330 & 0.13 & 1.72 & 0.20 \\
& 20120910\_A1 & 219 & 1.7  & 2.4 & 0.047 & 0.221 & 0.19 & 2.25 & 0.27 \\
& 20120917\_A1 &  192 & 0.36 & 2.2 & 0.043 & 0.219 & 0.12 & 2.09 & 0.14\\
& 20120917\_A3 & 607 & 4.23  & 2.2 & 0.033 & 0.259 & 0.10 & 5.57 & 0.15 \\
& 20120918\_A2 &  69 & 15.8 & 1.7 & 0.060 & 0.174 & 0.15 & 0.64 & 0.23 \\
& 20120918\_A3 & 214 & 0.89 & 1.7 & 0.041 & 0.230 & 0.20 & 2.04 & 0.36 \\
\hline
\multirow{7}{*}{\begin{minipage}[c]{3.5cm}\it Dasyhelea flavifrons  \\ \rm (Diptera: Ceratopogonidae) \\  \end{minipage}}
& 20110511\_A2 & 279 & 0.9  & 2.3 & 0.053 & 0.248 & 0.20 & 1.25 & 0.35 \\
& 20120702\_A1 &  98 & 2.1  & 2.0 &0.062 & 0.162 & 0.14 & 0.69 & 0.20 \\
& 20120702\_A2 & 111 & 7.3  & 2.0 & 0.056 & 0.169 & 0.13 & 0.88 & 0.18 \\
& 20120702\_A3 &  80 & 10.0 & 2.0 & 0.060 & 0.170 & 0.12 & 0.32 & 0.20 \\
& 20120703\_A2 & 167 & 4.4  & 1.8 & 0.046 & 0.140 & 0.07 & 0.52 & 0.12 \\
& 20120704\_A1 & 152 & 10.0 & 1.7 & 0.050 & 0.154 & 0.09 & 0.63 & 0.15 \\
& 20120704\_A2 & 154 & 5.3  & 1.7 & 0.053 & 0.160 & 0.08 & 0.61 & 0.13 \\
& 20120705\_A1 & 188 & 5.9  & 1.8 & 0.055 & 0.182 & 0.12 & 0.92 & 0.20 \\
\hline
\hline
\end{tabular}
\caption{
{\bf Swarm data.} 
Each line represents a different swarming event (acquisition). $N$ is the number of individuals in the swarm, $r_1$ the time average nearest neighbor distance in the particular acquisition, $r_0$ the average correlation length, $|\vec v|$ the average speed of the individuals, $l$ the body length, $\chi$ the average susceptibility and $\phi$ the average polarization. The average susceptibility in a system of noninteracting particles (with every quantity normalized as in natural swarms) is $\chi \sim 0.1$.}
\label{tavolone}
\end{table*}

\section*{Legend for Supplementary Video}

{\bf SM-Video1:}
Three dimensional visualization of a wild swarm of roughly $200$ midges in the field (Diptera:Chironomidae).
The swarm has been video recorded at 170 frames per seconds, with a resolution
of 4Mpx, by a IDT-M5 camera. This 3d reconstruction has been obtained through the dynamical
tracking algorithm based on our trifocal experimental technique.

\end{document}